\documentclass[a4paper, 10pt, conference]{ieeeconf}      

\IEEEoverridecommandlockouts                              
\usepackage{amsfonts}
\usepackage{amsmath} 
\usepackage{graphicx}

\newtheorem{theorem}{Theorem}[section]
\newtheorem{lemma}[theorem]{Lemma}
\newtheorem{definition}[theorem]{Definition}

\newtheorem{assumption}[theorem]{Assumptions}

\newcommand{\sr}{\stackrel}

\newcommand{\tri}{\sr{\triangle}{=}}

\newcommand{\noi}{\noindent}

\newcommand{\be}{\begin{equation}}
\newcommand{\ee}{\end{equation}}
\newcommand{\bea}{\begin{eqnarray}}
\newcommand{\eea}{\end{eqnarray}}
\newcommand{\bes}{\begin{eqnarray*}}
\newcommand{\ees}{\end{eqnarray*}}

\newcommand{\bfi}{\begin{figure}}
\newcommand{\bfit}{\begin{figure}[t]}
\newcommand{\bfib}{\begin{figure}[b]}
\newcommand{\bfih}{\begin{figure}[h]}
\newcommand{\bfip}{\begin{figure}[p]}
\newcommand{\efi}{\end{figure}}

\newcommand{\bi}{\begin{itemize}}
\newcommand{\ei}{\end{itemize}}
\newcommand{\ben}{\begin{enumerate}}
\newcommand{\een}{\end{enumerate}}

\title{\LARGE \bf
Causal Rate Distortion Function on Abstract Alphabets and Optimal Reconstruction Kernel}

\author{Charalambos D. Charalambous, Photios A. Stavrou and Christos K. Kourtellaris
\thanks{C. D. Charalambous (\tt\small chadcha@ucy.ac.cy). }%
\thanks{P. A. Stavrou  (\tt\small  stavrou.fotios@ucy.ac.cy). }%
\thanks{C. K. Kourtellaris  (\tt\small  kourtellaris.christos@ucy.ac.cy). }%
\thanks{The authors are  with the Department of Electrical and Computer Engineering,
University of Cyprus, Nicosia, CYPRUS }
}

\begin{document}

\maketitle
\thispagestyle{empty}
\pagestyle{empty}

\begin{abstract}
A Causal rate distortion function with a general fidelity criterion is formulated on abstract alphabets and the optimal reconstruction kernel is derived, which consists of a product of causal kernels. In the process, general abstract spaces are introduced to show existence of the minimizing kernel using weak$^*$-convergence. Certain properties of the causal rate distortion function are presented.
\end{abstract}

\section{INTRODUCTION}

\noi This paper is concerned with lossy data compression subject to distortion or fidelity criterion and causal decoding on abstract alphabets. Its information theoretic interpretation  is the causal rate distortion function  formulated  via the directed information between the source sequence $X^n\tri\{X_0,X_1,\ldots,X_n\}$ and its reproduction sequence $Y^n\tri\{Y_0,Y_1,\ldots,Y_n\}$ defined by
\begin{align}
I(X^n{\rightarrow}Y^n)&\tri \sum_{i=0}^{n}I(X^i; Y_i | Y^{i-1})\label{exis1}
\end{align}
The average distortion constraint is
\begin{align} 
E\{d_{0,n}(X^n,Y^n)\}\leq{D},\:d_{0,n}(x^n,y^n)\tri\sum^n_{i=0}\rho_{0,i}(x^i,y^i) \label{exis2}
\end{align}
where $D\geq0$,  $d_{0,n}(\cdot,\cdot)$  a non-negative distortion function. \\
Define the causal product of conditional distributions by
\begin{align}
{\overrightarrow P}_{Y^n|X^n}(dy^n|x^n)\tri\otimes^n_{i=0}P_{Y_i|Y^{i-1},X^i}(dy_i|y^{i-1},x^i)\label{exis3}
\end{align}
where $P_{Y_i|Y^{i-1},X^i}(dy_i|y^{i-1},x^i)$ denotes the conditional distribution of $Y_i$ given $(Y^{i-1},X^{i}),~i=0,1,\ldots,n.$\\
Since causal codes as defined in \cite{neu1982} satisfy $P_{X_i|X^{i-1},Y^{i-1}}(dx_i|x^{i-1},y^{i-1})=P_{X_i|X^{i-1}}(dx_i|x^{i-1})$.~$P-a.s$ (see also Lemma~\ref{lem1}), in the analysis it is convenient   to express $I(X^n\rightarrow{Y^n})$  as a functional of ${\overrightarrow P}_{Y^n|X^n}(dy^n|x^n)$ as follows.
\begin{align}
I(X^n\rightarrow{Y^n}) &=\int \log(\frac{{\overrightarrow P}_{Y^n|X^n}(dy^n|x^n)}{{P}_{Y^n}(dy^n)}) \nonumber \\
& \times {\overrightarrow P}_{Y^n|X^n}(dy^n|x^n)  P_{X^n}(dx^n) \\
& ={\mathbb I}(P_{X^n},{\overrightarrow P}_{Y^n|X^n})
\end{align}
where ${\mathbb I}(P_{X^n},{\overrightarrow P}_{Y^n|X^n})$ indicates the functional dependence of $I(X^n\rightarrow{Y^n})$ on $\{P_{X^n},{\overrightarrow P}_{Y^n|X^n}\}$.\\
The causal information rate distortion function investigated is 
\begin{align}
\inf_{{\overrightarrow P}_{Y^n|X^n}(dy^n|x^n):E\big\{d_{0,n}(X^n,Y^n)\big\}\leq D} I(X^n\rightarrow{Y^n})\label{exis4}
\end{align}
Under appropriate assumptions on $d_{0,n}(\cdot,\cdot)$ it is shown that the optimal causal product (reproduction channel) ${\overrightarrow P}^*_{Y^n|X^n}$ which achieves the infimum in (\ref{exis4}) is given by
\begin{align}
{\overrightarrow P}^*_{Y^n|X^n}(dy^n|x^n)=\otimes^n_{i=0}\frac{e^{s\rho_i(x^i,y^i)}P^*_{Y_i|Y_{i-1}}(dy_i|y^{i-1})}{\int_{{\cal Y}_i}e^{\rho_i(x^i,y^i)}P^*_{Y_i|Y_{i-1}}(dy_i|y^{i-1})} \label{sk}
\end{align}
where $s\leq0$ is the Lagrange multiplier associated with the fidelity constraint. The operational meaning of (\ref{exis4}) is shown in \cite{tatikonda00} via coding theorems (called sequential code), hence this aspect will not be discussed. Rather, the main emphasis  of the paper is the mathematical formulation, the prove of existence of solution to (\ref{exis4}), the derivation of (\ref{sk}), the derivation of a closed form  expression for the causal rate distortion function, and some of  its properties.   \\
The Shannon source code consists of an encoder-decoder pair. The encoder observes a source sequence $X^{\infty}\tri\{X_0,X_1,\ldots\}$ and generates a compressed representation  $\{Z_0,Z_1,\ldots\}$. The decoder upon observing the representation sequence $\{Z_0,Z_1,\ldots\}$ generates a reproduction sequence $Y_i=f_i(X^{\infty})$ of $X_i$, for every time step $i$. The dependence of the reproduction sequence on the future source symbols, in addition to its past and present symbols makes such a decoder non-causal. In Neuhoff and Gilbert \cite{neu1982}, a source code is defined as causal if the reproduction sequence is such that $f_i(X^{\infty})=f_i(\tilde X^{\infty})$ whenever $X^i={\tilde X}^i,~\forall i=0,1,\ldots$. The definition of a causal code necessitates that any  information theoretic causal rate distortion function should lead to an optimal reconstruction conditional distribution which is causally dependent on the source symbols, and (\ref{sk}) has this property.\\
The classical rate distortion function is defined via the mutual information between $X^n$ and $Y^n$, namely, $I(X^n;Y^n)$ with average distortion (\ref{exis2}), and  the code is assumed non-causal, leading to the well known optimal reconstruction  \cite{berger71,csiszar74}
\bea
P_{Y^n|X^n}^*(dy^n|x^n)=\frac{e^{   s\sum_{i=0}^n \rho_{0,i}(x^i,y^i) } P_{Y^n}^*(dy^n)}{\int_{{\cal Y}_{0,n}}e^{ s  \sum_{i=0}^n \rho_{0,i}(x^i,y^i)  }
P_{Y^n}^*(dy^n)} \label{eq.i.1}
\eea
Since by chain rule $P_{Y^n|X^n}(dy^n| X^n=x^n) = \otimes_{i=0}^n P_{Y_i|Y^{i-1},X^n=x^n}(dy_i|y^{i-1}=y^{i-1}, X^n=x^n)$, the classical rate distortion theory gives a reconstruction  $Y_i=y_i$ which depends on future values of the source symbols, $(X_{i+1}=x_{i+1}, \ldots, X_n=x_n)$ in addition to its past reconstructions  $Y^{i-1}=y^{i-1}$, and past and present source symbols $X^i=x^i$. The point to be made here is that, in general, aside from some special examples, such as the i.i.d source and single letter distortion $d_{0,n}=\sum^n_{i=0}\rho_i(x_i,y_i)$ \cite{thomas-cover91} the reconstruction conditional distribution and hence the decoder of the classical rate distortion function is non-causal. On the other hand, a code is causal if the reconstruction distribution is causal.

\section{PROBLEM FORMULATION}
In this section, we introduce the set up of the problem on discrete time sets $\mathbb{N}^n\tri\{0,1,\ldots,n\}$, $n \in \mathbb{N} \tri \{0,1,2,\ldots\}$. Assume all processes are defined on  a complete probability space
$(\Omega, {\cal F}(\Omega), \mathbb{P})$ with filtration $\{{\cal F}_t\}_{t
\geq 0}$. The source and reconstruction alphabets are sequences of Polish spaces \cite{dupuis-ellis97} $\{ {\cal X}_t: t\in\mathbb{N}\}$ and $\{ {\cal Y}_t: t\in\mathbb{N}\}$, respectively, (e.g., ${\cal Y}_t, {\cal X}_t$ are complete separable metric spaces), associated with their corresponding measurable spaces $({\cal X}_t,{\cal B}({\cal X}_t))$ and $({\cal Y}_t, {\cal B}({\cal Y}_t))$  (e.g., ${\cal B}({\cal X}_t)$ is a Borel $\sigma-$algebra of subsets
of the set ${\cal X}_t$ generated by closed sets), $t\in\mathbb{N}$. Sequences of alphabets are  identified
with the product spaces $({\cal X}_{0,n},{\cal B}({\cal X}_{0,n})) \tri  \times_{k=0}^{n}({\cal X}_k,{\cal B}({\cal X}_k))$,
and $({\cal Y}_{0,n},{\cal B}({\cal Y}_{0,n}))\tri \times_{k=0}^{n}({\cal Y}_k,{\cal B}({\cal Y}_k))$.
The source and reconstruction are processes denoted by $X^n \tri \{X_t: t\in\mathbb{N}^n\}$, $X:\mathbb{N}^{n}\times\Omega\mapsto {\cal X}_t$, and by $Y^n\tri \{Y_t: t\in\mathbb{N}^n\}$, $Y:\mathbb{N}^{n}\times\Omega\mapsto  {\cal Y}_t$, respectively. Probability measures on any measurable space  $( {\cal Z}, {\cal B}({\cal Z}))$ are denoted by ${\cal M}_1({\cal Z})$. It is assumed  that the $\sigma$-algebras $\sigma\{X^{-1}\}=\sigma\{Y^{-1}\}=\{\emptyset,\Omega\}$.

\begin{definition}\label{stochastic kernel}
Let $({\cal X}, {\cal B}({\cal X})), ({\cal Y}, {\cal B}({\cal Y}))$ be measurable spaces in which $\cal Y$ is a Polish Space.\\
A  stochastic Kernel on $\cal Y$ given $\cal X$ is a mapping $q: {\cal B}({\cal Y}) \times {\cal X}  \rightarrow [0,1]$ satisfying the following two properties:
\par 1) For every $x \in {\cal X}$, the set function $q(\cdot;x)$ is a probability measure (possibly finitely additive) on ${\cal B}({\cal Y}).$
\par 2) For every $F \in {\cal B}({\cal Y})$, the function $q(F;\cdot)$ is ${\cal B}({\cal X})$-measurable.\\
 The set of all such stochastic Kernels is denoted by ${\cal Q}({\cal Y};{\cal X})$.
\end{definition}
\par An important notion is conditional independence.
The Random Variable (R.V.) ${Z}$ is called conditional independent of R.V. $X$ given the R.V. $Y$ if and only if $X\leftrightarrow Y \leftrightarrow Z$ forms a Markov chain in both directions.
\par Stochastic kernels can be used to define non-causal and causal product reconstruction kernels and associated rate distortion functions.
\begin{definition}\label{comprchan}
Given measurable spaces $({\cal X}_{0,n},{\cal B}({\cal X}_{0,n}))$, $({\cal Y}_{0,n},{\cal B}({\cal Y}_{0,n}))$, and their product spaces, data compression channels are defined as follows.
\begin{enumerate}
\item {\it A Non-Causal Data Compression Channel} is a  stochastic kernel $ q_{0,n} (dy^n; x^n) \in {\cal Q}({\cal Y}_{0,n} ;{\cal X}_{0,n}), n \in \mathbb{N}$.
\item{\it A Causal Product Data Compression Channel} is a product of a sequence of causal stochastic kernels defined by
\bes
{\overrightarrow q}_{0,n}(dy^n;x^n)&=\otimes_{i=0}^n q_i(dy_i;y^{i-1},x^i)
\ees
\end{enumerate}
where $q_i \in {\cal Q}({\cal Y}_i;{\cal Y}_{0,i-1}\times{\cal X}_{0,i}), i=0,\ldots,n,~n \in \mathbb{N}$.
\end{definition}
\noi Note that classical rate distortion theory is concerned with finding the optimal $P_{Y^n|X^n}(dy^n| X^n=x^n)$, which is generally non-causal, while in this paper the interest is to find the optimal causal product kernel. 
\subsection{Causal and Classical Rate Distortion Functions}
In this section the classical rate distortion function which has a non-causal structure is reviewed, and then the causal rate distortion function is defined.\\
Given a source probability measure ${\cal \mu}_{0,n} \in {\cal M}_1({\cal X}_{0, n})$ (possibly finite additive) and a reconstruction Kernel $q_{0,n} \in {\cal Q}({\cal Y}_{0, n};{\cal X}_{0, n})$, one can define three probability measures as follows.
\par (P1): The joint measure $P_{0,n} \in {\cal M}_1({\cal Y}_{0,n}\times {\cal X}_{0, n})$:
\begin{align}
P_{0,n}(G_{0,n})&\tri(\mu_{0,n} \otimes q_{0,n})(G_{0,n}),\:G_{0,n} \in {\cal B}({\cal X}_{0,n})\times{\cal B}({\cal Y}_{0,n})\nonumber\\
&=\int_{{\cal X}_{0,n}} q_{0,n}(G_{0,n,x^n};x^n) \mu_{0,n}(d{x^n})\nonumber
\end{align}
where $G_{0,n,x^n}$ is the $x^n-$section of $G_{0,n}$ at point ${x^n}$ defined by $G_{0,n,x^n}\tri \{y^n \in {\cal Y}_{0,n}: (x^n, y^n) \in G_{0,n}\}$ and $\otimes$ denotes the convolution.
\par (P2): The marginal measure $\nu_{0,n} \in {\cal M}_1({\cal Y}_{0,n})$:
\begin{align}
\nu_{0,n}(F_{0,n})&\tri P_{0,n}({\cal X}_{0, n} \times F_{0,n}),~F_{0,n} \in {\cal B}({\cal Y}_{0,n})\nonumber\\
&=\int_{{\cal X}_{0, n}} q_{0,n}(({\cal X}_{0, n}\times F_{0,n})_{{x}^{n}};{x}^{n}) \mu_{0,n}(d{x^n})\nonumber \\
&=\int_{{\cal X}_{0, n}} q_{0,n}(F_{0,n};x^n) \mu_{0,n}(dx^n)\nonumber
\end{align}
\par(P3): The product measure  $\pi_{0,n}:{\cal B}({\cal X}_{0,n}) \times
{\cal B}({\cal Y}_{0,n}) \mapsto [0,1] $ of $\mu_{0,n}\in{\cal M}_1({\cal X}_{0, n})$ and $\nu_{0,n}\in{\cal M}_1({\cal Y}_{0, n})$:
\begin{align}
\pi_{0,n}(G_{0,n})&\tri(\mu_{0,n} \times \nu_{0,n})(G_{0,n}),~G_{0,n} \in {\cal B}({\cal X}_{0,n}) \times {\cal B}({\cal Y}_{0,n})\nonumber\\
&=\int_{{\cal X}_{0, n}} \nu_{0,n}(G_{0,n,x^n}) \mu_{0,n}(dx^n)\nonumber
\end{align}
\noi The precise definition of mutual information between two sequences of Random Variables $X^n$ and $Y^n$, denoted $I(X^n; Y^n)$ is
defined via the Kullback-Leibler distance (or relative entropy) between the joint probability distribution of  $(X^n, Y^n)$ and the product of its marginal probability distributions of $X^n$ and $Y^n$, using the Radon-Nikodym derivative.
 Hence, by the construction of probability measures (P1)-(P3), and the chain rule of relative entropy \cite{dupuis-ellis97}:
\bea
&& I(X^n;Y^n) \tri  \mathbb{D}(P_{0,n}|| \pi_{0,n})\label{eq1}\\
&&=\int_{{\cal X}_{0,n} \times {\cal Y}_{0,n}}\log \Big( \frac{d  (\mu_{0,n} \otimes q_{0,n}) }{d ( \mu_{0,n} \times \nu_{0,n} ) }\Big) d(\mu_{0,n} \otimes q_{0,n}) \nonumber\\
&& = \int_{{\cal X}_{0,n} \times {\cal Y}_{0,n}} \log \Big( \frac{q_{0,n}(d y^n; x^n)}{  \nu_{0,n} (dy^n)   } \Big)\nonumber\\
&&q_{0,n}(dy^n;dx^n) \mu_{0,n}(dx^n) \nonumber\\
&&=\int_{{\cal X}_{0,n}} \mathbb{D}(q_{0,n}(\cdot;x^n)|| \nu_{0,n}(\cdot)) \mu_{0,n}(dx^n)\nonumber\\
&&\equiv \mathbb{I}(\mu_{0,n}; q_{0,n})  \label{re3}
 \eea
 Note that $(\ref{re3})$ states that mutual information is expressed as a functional of $\{\mu_{0,n}, q_{0,n}\}$ and it is denoted by $\mathbb{I}(\mu_{0,n}; q_{0,n})$. Note that necessary and sufficient conditions for existence of  a Radon-Nikodym derivative  for finitely additive measures can be found in \cite{maynard79}.
Moreover, $I(X^n;Y^n)$ is also expressed by the sum of two directed information as follows
\begin{align}
I(X^n;Y^n)&=I(X^n{\rightarrow}Y^n)+I(X^n{\leftarrow}Y^n)
\end{align}
where
\begin{align}
I(X^n{\rightarrow}Y^n)&\tri \sum_{i=0}^{n}I(X^i; Y_i | Y^{i-1})\\
I(X^n{\leftarrow}Y^n)&\tri \sum_{i=0}^{n}I(Y^{i-1};X_i | X^{i-1})
\end{align}
\begin{definition}\label{2.4}
(Classical Rate Distortion Function) Let $d_{0,n}: {\cal X}_{0,n}  \times {\cal Y}_{0,n} \rightarrow [0, \infty)$, be an ${\cal B}({\cal X}_{0,n}) \times {\cal B }( {\cal Y}_{0,n})$-measurable distortion function, and let $Q_{0,n}(D) \subset {\cal Q}({\cal Y}_{0,n}; {\cal X}_{0,n})$ (assuming is non-empty) denotes the average distortion or fidelity constraint defined by
\bea
&&Q_{0,n}(D)\tri \Big\{ q_{0,n} \in {\cal Q}({\cal Y}_{0,n}; {\cal X}_{0,n}):\nonumber\\
&&\frac{1}{n+1}\int_{{\cal X}_{0,n}}\int_{ {\cal Y}_{0,n}} d_{0,n}(x^n,y^n) q_{0,n}(dy^n;x^n)\nonumber\\
&&\mu_{0,n}(dx^n) \leq D \Big\},~D\geq 0 \label{dc1}
\eea
The classical rate distortion function associated with the non-causal kernel $q_{0,n} \in {\cal Q}({\cal Y}_{0,n}; {\cal X}_{0,n})$ is defined by
\bea
R_{0,n}(D) \tri \inf_{q_{0,n} \in Q_{0,n}(D)}\frac{1}{n+1} \mathbb{I}(\mu_{0,n};q_{0,n})  \label{f3s}
\eea
while its operational meaning can be established via ${\lim}{\sup_{n\rightarrow\infty}}{R_{0,n}}$.
\end{definition}
Existence in (\ref{f3s}) is shown assuming $d_{0,n}(x^n;\cdot)$ is bounded continuous on ${\cal Y}_{0,n}$ and ${\cal Y}_{0,n}$ is compact, using weak-convergence of probability measures in \cite{csiszar74}, and for more general $d_{0,n}(x^n;\cdot)$ which is only continuous in ${\cal Y}_{0,n}$ using weak*-convergence of measures \cite{farzad06} on Polish spaces. \\
A version of the optimal reconstruction kernel which attains the infimum in (\ref{f3s}), \cite{csiszar74} is
\bea
q_{0,n}^*(dy^n; x^n)=\frac{\ e^{s d_{0,n}(x^n,y^n)} \nu_{0,n}^*(dy^n) }{\int_{{\cal Y  }_{0,n}
}e^{s d_{0,n}(x^n,y^n)} \nu_{0,n}^*(dy^n)}, \quad  s \leq 0 \label{f6a}
\eea
where $\nu_{0,n}^* \in {\cal M}_1({\cal Y}_{0,n})$ is the marginal of $P_{0,n}^*= \mu_{0,n}\otimes q_{0,n}^* \in {\cal M}_1({\cal X}_{0,n}\times{\cal Y}_{0,n})$ and $s\leq 0$ is the Lagrange multiplier associated with the fidelity constraint $(\ref{dc1})$. Unfortunately, for general sources and distortion function $d_{0,n}$, the optimal reconstruction $q^*_{0,n}(dy^n;x^n)=\otimes^n_{i=0}q^*_i(dy_i;y^{i-1},x^n)$ is non-causal and introduces delay in the reconstruction processes. On the other hand, if the solution (\ref{f6a}) gives a reconstruction such that $q^*_{0,n}(dy^n;x^n)={\overrightarrow q}^*_{0,n}(dy^n;x^n)=\otimes^n_{i=0}q^*_i(dy_i;y^{i-1},x^i)$ it will be causal. However, there are only limited examples in which $(\ref{f6a})$ is causal on the source sequence. For single letter distortion function $d_{0,n}(x^n,y^n)=\frac{1}{n+1}\sum^n_{i=0}\rho_i(x_i,y_i)$ and independent sources $\mu_{0,n}(dx^n)=\otimes^n_{i=0}\mu_i(dx^i)$ (e.g., $\{X_i: i\in\mathbb{N}\}$ are independent) the optimal reconstruction $q^*_{0,n}(dy^n;x^n)$ factors into a product of causal kernels $q^*_{0,n}(dy^n;x^n)=\otimes^n_{i=0}q_i(dy_i,x_i)$ \cite{thomas-cover91}. This raises the question whether the classical rate distortion function can be reformulated using the causal product ${\overrightarrow q}_{0,n}(dy^n;x^n)$.
\par The next lemma relates causal product reconstruction kernels, mutual information, directed information, and conditional independence.
\begin{lemma} \label{lem1}
The following are equivalent for each $n\in\mathbb{N}$.
\begin{enumerate}
\item $q_{0,n} (dy^n; x^n)={\overrightarrow q}_{0,n}(dy^n;x^n)$, as defined in Definition \ref{comprchan}-2)

\item For each $i=0,1,\ldots, n-1$,  $Y_i \leftrightarrow (X^i, Y^{i-1}) \leftrightarrow (X_{i+1}, X_{i+2}, \ldots, X_n)$, forms a Markov chain

\item $I(X^n ; Y^n)=I(X^n \rightarrow Y^n)$

\item $I(X^n\leftarrow Y^n)=0$

\item For each  $i=0,1,\ldots, n-1$, $Y^i \leftrightarrow X^i \leftrightarrow X_{i+1}$ forms a Markov chain
\end{enumerate}
\end{lemma}
\noi{Proof.} Omitted due to space limitation.\\
According to Lemma~\ref{lem1} any source with a satisfying conditional distribution $P_{X_i|X^{i-1},Y^{i-1}}(dx_i|X^{i-1}=x^{i-1},Y^{i-1}=y^{i-1})=P_{X_i|X^{i-1}}(dx_i|X^{i-1}=x^{i-1}),~P-a.s.,$ $\forall{i}\in \mathbb{N}$ is equivalent to any of the equivalent statements of Lemma~\ref{lem1}. Therefore, for such a source the mutual information becomes
\bea
&&I(X^n; Y^n)=I(X^n{\rightarrow}Y^n)\nonumber\\
&&=\int_{{\cal X}_{0,n} \times {\cal Y}_{0,n}   } \log \Big( \frac{ \overrightarrow{q}_{0,n}(d y^n; x^n)}{  \nu_{0,n} (dy^n)   } \Big)\nonumber\\
&&\overrightarrow{ q}_{0,n}(dy^n;dx^n)\mu_{0,n}(dx^n) \label{ex10}  \\
&&\equiv {\mathbb I}(\mu_{0,n}; \overrightarrow{q}_{0,n} )  \label{ex11}
\eea
where (\ref{ex11}) states that $I(X^n;Y^n)$ is a functional of $\{\mu_{0,n},{\overrightarrow q}_{0,n}\}$.
Hence, causal rate distortion is defined by optimizing ${\mathbb I}(\mu_{0,n}; \overrightarrow{q}_{0,n})$ over ${\overrightarrow q}_{0,n}$ which satisfies a distortion constraint.


\begin{definition}\label{def1}
(Causal Rate Distortion Function)
Suppose $d_{0,n}\tri\sum^n_{i=0}\rho_{0,i}(x^i,y^i)$, where $\rho_{0,i}: {\cal X}_{0,i}  \times {\cal Y}_{0,i}\rightarrow [0, \infty)$, is a sequence of ${\cal B}({\cal X}_{0,i}) \times {\cal B }( {\cal Y}_{0,i})$-measurable distortion functions, and let $\overrightarrow{Q}_{0,n}(D)$ (assuming is non-empty) denotes the average distortion or fidelity constraint defined by
\bea
&&\overrightarrow{Q}_{0,n}(D)\tri\Big\{\overrightarrow{q}_{0,i} \in  {\cal M}_1({\cal Y}_{0,i}),0\leq{i}\leq{n}:\nonumber\\
&&\frac{1}{n+1}\sum_{i=0}^{n}\int_{{\cal X}_{0,i}}\int_{{{\cal Y}}_{0,i}}\rho_{0,i}({x^i},{y^i})\overrightarrow{q}_{0,i}(d{y}^{i};{x}^{i})\nonumber\\
&&\mu_{0,i}(d{x}^i)\leq D\Big\},~D\geq0 \label{eq2}
\eea
The causal rate distortion function associated with the causal product kernel ${\overrightarrow q}_{0,n} \in {\overrightarrow Q}_{0,n}(D)$ is defined by
\bea
{\overrightarrow R}_{0,n}(D) \tri  \inf_{{\overrightarrow{q}_{0,n}\in \overrightarrow{Q}_{0,n}(D)}}\frac{1}{n+1}{\mathbb I}(\mu_{0,n};\overrightarrow{q}_{0,n})
\label{ex12}
\eea
while its operational meaning can be established via $\lim\sup_{n\rightarrow{\infty}}{\overrightarrow R}_{0,n}$.
\end{definition}
Clearly, ${\overrightarrow R}_{0,n}(D)$ is characterized by minimizing directed information or equivalently $\mathbb{I}(\mu_{0,n};\overrightarrow{q}_{0,n})$ over the causal product measure ${\overrightarrow q}_{0,n}\in{\overrightarrow Q}_{0,n}(D)$.
\begin{lemma}\label{th1}
$\overrightarrow{q}_{0,n}\in {\cal M}_1({\cal Y}_{0,n})$ is uniquely determined by $\{q_{i}\in {\cal Q}_{i}({\cal Y}_{i} ; {\cal Y}_{0,i-1} \times{\cal X}_{0,i}) \}_{i=0}^n$ and vice-versa, $P-a.s$.
\end{lemma}
{Proof.} For densities this result is derived in \cite{Perm}.

\section{EXISTENCE OF OPTIMAL CAUSAL PRODUCT RECONSTRUCTION KERNEL}
\par In this section, appropriate topologies and function spaces are employed to show existence of the minimizing causal product kernel in $(\ref{ex12})$. In the process we also show existence for $R_{0,n}(D)$.

\subsection{Abstract Spaces}

\par Let $BC({\cal Y}_{0,n})$ denote the vector space of bounded continuous real valued functions defined
on the Polish space ${\cal Y}_{0,n}$. Furnished with the sup norm
topology, this is a Banach space. The topological dual of $BC({\cal Y}_{0,n})$ denoted by $ \Big( BC({\cal Y}_{0,n})\Big)^*$ is isometrically isomorphic to the Banach space of finitely additive regular bounded signed measures on ${\cal Y}_{0,n}$ \cite{dunford-schwartz58}, denoted by $M_{rba}({\cal Y}_{0,n})$. Let $\Pi_{rba}({\cal Y}_{0,n})\subset M_{rba}({\cal Y}_{0,n})$ denote the set of regular bounded
finitely additive probability measures on ${\cal Y}_{0,n}$.  Clearly if ${\cal Y}_{0,n}$ is compact,
then $\Big(BC({\cal Y}_{0,n})\Big)^*$ will be isometrically isomorphic to the space of countably additive signed measures, as
in \cite{csiszar74}. Denote by $L_1(\mu_{0,n}, BC({\cal Y}_{0,n}))$ the space of all $\mu_{0,n}$-integrable functions defined on  ${\cal X}_{0,n}$ with values in $BC({\cal Y}_{0,n}),$ so that for each $\phi \in L_1(\mu_{0,n}, BC({\cal Y}_{0,n}))$ its norm is defined by
\bes
\parallel \phi \parallel_{\mu_{0,n}} \tri \int_{{\cal X}_{0,n}} ||\phi(x^n)(\cdot) ||_{BC({\cal Y}_{0,n})} \mu_{0,n}(dx^n) <
\infty
\ees
The norm topology $\parallel{\phi}\parallel_{\mu_{0,n}}$, makes $L_1(\mu_{0,n}, BC({\cal Y}_{0,n}))$ a Banach
space, and it follows from the theory of ``lifting" \cite{aitulcea69} that the dual
of this space is $L_{\infty}^w(\mu_{0,n}, M_{rba}({\cal Y}_{0,n}))$, denoting
the space of all $M_{rba}({\cal Y}_{0,n})$ valued functions $\{q\}$ which
are weak$^*$-measurable in the sense that for each $\phi \in
BC({\cal Y}_{0,n}),$  $x^n \longrightarrow q_{x^n}(\phi) \tri \int_{{\cal Y}_{0,n}}\phi(y^n) q(dy^n;x^n)$ is $\mu_{0,n}$-measurable and $\mu_{0,n}$-essentially
bounded.
\subsection{Weak$^*$-Compactness and Existence}
Define an admissible set of stochastic kernels associated with classical rate distortion function by
\bes
Q_{ad}\tri L_{\infty}^w(\mu_{0,n}, \Pi_{rba}({\cal Y}_{0,n})) \subset L_{\infty}^w(\mu_{0,n}, M_{rba}({\cal Y}_{0,n}))
\ees
Clearly, $Q_{ad}$ is a unit sphere in $L_{\infty}^w(\mu_{0,n}, M_{rba}({\cal Y}_{0,n}))$. For each $\phi{\in}L_1(\mu_{0,n}, BC({\cal Y}_{0,n}))$ we can define a linear functional on $L_{\infty}^w(\mu_{0,n}, M_{rba}({\cal Y}_{0,n}))$ by
\bes
\ell_{\phi}(q_{0,n})\tri\frac{1}{n+1}\int_{{\cal X}_{0,n}}\Big( \int_{{\cal Y}_{0,n}} \phi(x^n,y^n)\\
q_{0,n}(dy^n;x^n) \Big)\mu_{0,n}(dx^n)
\ees
This is a bounded, linear and weak$^*$-continuous functional on $L_{\infty}^w(\mu_{0,n}, M_{rba}({\cal Y}_{0,n}))$. For $d_{0,n}: {\cal X}_{0,n}  \times {\cal Y}_{0,n}\rightarrow [0, \infty)$ measurable and $d_{0,n}{\in}L_1(\mu_{0,n},BC({\cal Y}_{0,n}))$ the distortion constraint set of the classical rate distortion function is
\bes
Q_{0,n}(D)\tri\{q{\in}Q_{ad}:\frac{1}{n+1}\ell_{d_{0,n}}(q_{0,n}){\leq}D\}
\ees
%
It can be shown that $Q_{0,n}(D)$ is bounded and weak$^*$-closed subset of $Q_{ad}$ and hence weak$^*$-compact (Compactness of $Q_{ad}$ follows from Alaoglu's Theorem~\cite {dunford-schwartz58},\cite{rudin91}).\\
Next, we define the set of causal product kernels as follows.
\begin{align}
{\overrightarrow \Pi}_{rba}({\cal Y}_{0,n})&=\Big\{{\overrightarrow q}_{0,n}(dy^n;x^n)\tri\otimes_{i=1}^n {q}_i(dy_i;y^{i-1},x^i):\nonumber\\
&{q}_i(dy_i;y^{i-1},x^i)\in{\Pi}_{rba}({\cal Y}_{i}), \: i\in\mathbb{N}^n \Big\} \nonumber
\end{align}
where $L_{\infty}^w(\mu_{0,n}, {\overrightarrow \Pi}_{rba}({\cal Y}_{0,n}))$ denotes the space of all ${\overrightarrow \Pi}_{rba}({\cal Y}_{0,n})$ valued functions $\{\overrightarrow q\}$ which are weak$^*$-measurable in the sense that for each $\phi \in
BC({\cal Y}_{0,n}),$  $x^n \rightarrow {\overrightarrow q}_{x^n}(\phi) \tri \int_{{\cal Y}_{0,n}}\phi(y^n) {\overrightarrow q}(dy^n;x^n)$ is $\mu_{0,n}$-measurable and $\mu_{0,n}$-essentially bounded.\\
Define the admissible set of causal product stochastic kernels associated with the causal rate distortion function by
\begin{align}
{\overrightarrow Q}_{ad}&\tri L_{\infty}^w(\mu_{0,n}, {\overrightarrow \Pi}_{rba}({\cal Y}_{0,n}))\nonumber
\end{align}
Clearly, ${\overrightarrow Q}_{ad}=\{q_{0,n} \in Q_{ad}:q_{0,n}(dy^n;x^n)=\overrightarrow{q}_{0,n}(dy^n; x^n)\}$.
For $d_{0,n}: {\cal X}_{0,n}  \times {\cal Y}_{0,n}\rightarrow [0, \infty)$ which is measurable and $d_{0,n}{\in}L_1(\mu_{0,n},BC({\cal Y}_{0,n}))$ the distortion constraint of causal rate distortion function is 
\bes
&&{\overrightarrow Q_{{0,n}}(D)} \tri \Big\{{\overrightarrow q}_{0,n} \in {\overrightarrow Q}_{ad} :\\
&&\frac{1}{n+1}\ell_{d_{0,n}}({\overrightarrow q}_{0,n})\tri \int_{{\cal X}_{0,n}} \biggr(\int_{{\cal Y}_{0,n}}d_{0,n}(x^n,y^n)\\
&&{\overrightarrow q}_{0,n}(dy^n;x^n) \biggr)\mu_{0,n}(dx^n)\leq D\Big\}
\ees
\begin{assumption}\label{as1}
\noi We make the following assumptions.
\begin{enumerate}
\item The set $\overrightarrow Q_{ad}$ is weak$^*$-closed.
\item The set ${\overrightarrow Q_{0,n}(D)}$ is non-empty.
\end{enumerate}
\end{assumption}
\begin{lemma} \label{wkstr}
Suppose Assumptions~\ref{as1} hold. Let  ${\cal X}_{0,n}, {\cal Y}_{0,n}$ be two Polish spaces  and $d_{0,n} : {\cal X}_{0,n}\times {\cal Y}_{0,n} \rightarrow
[0,\infty],$ a   measurable, non-negative, extended real valued function, such that $d_{0,n}{\in}L_1(\mu_{0,n},BC({\cal Y}_{0,n}))$.
For any $D \in [0,\infty)$, the set ${\overrightarrow Q_{0,n}(D)}$ is weak$^*$-compact.
\end{lemma}
{Proof.} By Assumptions~\ref{as1}, $\overrightarrow Q_{ad}$ is a weak$^*$-closed, hence as a subset of a weak$^*$-compact set $Q_{ad}$ it is weak$^*$-compact. Also, under assumptions~\ref{as1}, ${\overrightarrow Q_{0,n}(D)}$ is bounded and weak$^*$-closed and hence it is weak$^*$-compact (as a weak$^*$-closed subset of the weak$^*$-compact set ${\overrightarrow Q}_{ad}$) $\bullet$
\begin{theorem} \label{th3}
Under Assumptions~\ref{as1}, ${\overrightarrow R}_{0,n}(D)$ has a minimum.
\end{theorem}
{Proof.} Follows from Lemma~\ref{wkstr} and the lower semi-continuity of ${\mathbb{I}}(\mu_{0,n};\cdot)$ on ${\overrightarrow Q}_{ad}$ $\bullet$



\section{NECESSARY CONDITIONS OF OPTIMALITY OF CAUSAL PRODUCT RATE DISTORTION FUNCTION}

In this section the form of the optimal causal product reconstruction kernels is derived. The method is based on calculus of variations on the space of measures \cite{dluenberger69}.
\begin{theorem} \label{th5}

Suppose ${\mathbb I}_{\mu_{0,n}}({\overrightarrow q}_{0,n}) \tri {\mathbb I}(\mu_{0,n}; \overrightarrow{q}_{0,n})$ is well defined for every ${\overrightarrow q}_{0,n}\in L_{\infty}^w({\mu_{0,n},\overrightarrow \Pi}_{rba}({\cal Y}_{0,n}))$ possibly taking values
from the set $[0,\infty].$ Then  ${\overrightarrow q}_{0,n} \rightarrow {\mathbb I}_{\mu_{0,n}}({\overrightarrow q}_{0,n})$ is
Gateaux differentiable at every point in $L_{\infty}^w({\mu_{0,n},\overrightarrow \Pi}_{rba}({\cal Y}_{0,n})),$  and the Gateaux
derivative at the  point ${\overrightarrow q}_{0,n}^0$ in the direction ${\overrightarrow q}_{0,n}-{\overrightarrow q}_{0,n}^0$ is given
by
\bes
&&\delta{\mathbb I}_{\mu_{0,n}}({\overrightarrow q}_{0,n}^0;{\overrightarrow q}_{0,n}-{\overrightarrow q}_{0,n}^0)\\
&&=\int_{{\cal X}_{0,n}}\int_{{\cal Y}_{0,n}}\log \Bigg(
\frac{{\overrightarrow q}_{0,n}^0(dy^n;x^n)}{\nu_{0,n}^0(dy^n)}\Bigg)\\
&& ({\overrightarrow q}_{0,n}-{\overrightarrow q}_{0,n}^0)(dy^n;x^n) \mu_{0,n}(dx^n)
\ees
where $\nu_{0,n}^0\in{\cal M}_1({\cal Y}_{0,n})$ is the marginal measure corresponding
to ${\overrightarrow q}_{0,n}^0\otimes\mu_{0,n}(dx^n)\in{\cal M}_1({\cal Y}_{0,n}\times{\cal X}_{0,n})$.
\end{theorem}
\noi {Proof.} The proof is based on the fact that the causal product stochastic kernel ${\overrightarrow q}_{0,n}$ is used to show the existence of Gateaux Differential \cite{dluenberger69} rather than for individual causal stochastic kernel $q_i(dy_i;y^{i-1},x^i)$, $i\in\mathbb{N}^n$ $\bullet$
\par The constrained problem defined by (\ref{ex12}) can be reformulated using Lagrange multipliers as follows (equivalence of constrained and unconstrained problems follows from \cite{dluenberger69}).
\bea
&&{\overrightarrow R}_{0,n}(D) = \inf_{{\overrightarrow q}_{0,n} \in {\overrightarrow Q}_{ad}} \Big\{ \frac{1}{n+1}{{\mathbb I}}(\mu_{0,n};{\overrightarrow q}_{0,n})\nonumber\\
&&-s(\ell_{{d}_{0,n}}({\overrightarrow q}_{0,n})-D)\Big\} \label{ex13}
\eea
and  $s \in(-\infty,0]$ is the Lagrange multiplier.
\begin{theorem} \label{th6}
Suppose $d_{0,n}(x^n,y^n)=\sum_{i=0}^n\rho_{0,i}(x^i,y^i)$ and the assumptions of Lemma~\ref{wkstr} hold. The infimum in $(\ref{ex13})$ is attained at  $\overrightarrow{q}^*_{0,n} \in
L_{\infty}^w(\mu_{0,n},{\overrightarrow \Pi}_{rba}({\cal Y}_{0,n}))$ given by
\bea
\overrightarrow{q}^*_{0,n}(dy^n;x^n)=\otimes_{i=0}^n\frac{e^{s \rho_i(x^i,y^i)}
\nu^*_i(dy^i;y^{i-1})}{\int_{{\cal Y}_i} e^{s \rho_i(x^i,y^i)} \nu^*_i(dy_i;y^{i-1})}\label{ex14}
\eea
and $\nu^*_i(dy_i;y^{i-1})\in {\cal Q}({\cal Y}_i;{\cal Y}_{0,{i-1}})$. The causal rate distortion
function is given by
\bea
&&{\overrightarrow R}_{0,n}(D)=sD - \frac{1}{n+1}\sum_{i=0}^n\int_{{{\cal X}_{0,i}}\times{{\cal Y}_{0,i-1}}}\nonumber\\
&& \log \Big( \int_{{\cal Y}_i} e^{s
\rho_i(x^i,y^i)} \nu^*_i(dy_i;y^{i-1})\Big)\nonumber\\
&&{{\overrightarrow q}^*_{0,i-1}}(dy^{i-1};x^{i-1})\otimes \mu_{0,i}(dx^i)\label{ex15}
\eea
If ${\overrightarrow R}_{0,n}(D) > 0$ then $ s < 0$  and
\bes
\frac{1}{n+1}\sum_{i=0}^n\int_{{\cal X}_{0,i}} \int_{{\cal Y}_{0,i}}
\rho_{0,i}(x^i,y^i) {\overrightarrow q}^*_{0,i}(dy^i;x^i) \mu_{0,i}(dx^i)=D
\ees
\end{theorem}
\noi {Proof.} The fully unconstraint problem of (\ref{ex13}) is obtained by introducing another Lagrange multiplier. Using this and Theorem~\ref{th5} we obtain (\ref{ex14}) and (\ref{ex15})  $\bullet$
\section{PROPERTIES OF CAUSAL RATE DISTORTION FUNCTION}
In this section, we present some important properties of the causal rate distortion function as it is defined in (\ref{ex12}).
\begin{theorem} \label{prop1}
\noi
\begin{enumerate}
\item ${\overrightarrow R}_{0,n}(D)$ is a convex, non-increasing function of $D$
\item If $\rho_{0,i} \in L^1(\pi_{0,i})$ then \\
a) ${\overrightarrow R}_{0,n}(\frac{1}{n+1}\sum_{i=0}^nE_{\pi_{0,i}}(\rho_{0,i}))=0$; \\
b) ${\overrightarrow R}_{0,n}(D)$ is non-increasing for $D \in [0, D_{max}]$ where $D_{max}=\frac{1}{n+1}\sum_{i=0}^nE_{\pi_{0,i}}(\rho_{0,i})$ and ${\overrightarrow R}_{0,n}(D)=0$ for any $D \geq D_{max}$
\item ${\overrightarrow R}_{0,n}(D) > 0$ for all $D < D_{max}$ and ${\overrightarrow R}_{0,n}(D)=0$ for all $D \geq D_{max}$, where
\bes
D_{max}= \min_{\{y^n\}\in {\cal Y}_{0,n}}\frac{1}{n+1}\sum_{i=0}^n \int_{{\cal X}_{0,i}} \rho_{0,i}(x^i,y^i)\mu_{0,i}(dx^i)
\ees
if such a minimum exists.
\end{enumerate}
\end{theorem}
\noi {Proof.} Omitted due to space limitation.

\section{CONCLUSION AND FUTURE WORK}

\subsection{Conclusion}
The solution of the causal rate distortion function subject to a reproduction kernel which is a product of causal kernels is presented, on abstract alphabets. Some of its properties are also presented. It is believed that the optimal reconstruction kernel as a product of causal kernels has several implications in applications where causality of the decoder as a function of the source is of concern.
\subsection{Future Work}
Examples are currently under investigation, and will be presented at the final version of the paper.

\section{APPENDIX}

\end{document}